\documentclass[twocolumn,showpacs,floatfix,preprintnumbers,prb]{revtex4-1}
\usepackage{graphicx}
\usepackage{float}
\usepackage{dcolumn}
\usepackage{bm}
\usepackage{amssymb}

\begin{document}

\title{Transport Properties and Optical Conductivity of the adiabatic Su-Schrieffer-Heeger model: a showcase study for rubrene based field effect transistors.}

\author{V. Cataudella}
\affiliation{CNR-SPIN and Dipartimento di Scienze Fisiche, Univ. di Napoli ``Federico
II'', I-80126 Italy}
\author{G. De Filippis}
\affiliation{CNR-SPIN and Dipartimento di Scienze Fisiche, Univ. di Napoli ``Federico
II'', I-80126 Italy}
\author{C.A. Perroni}
\affiliation{CNR-SPIN and Dipartimento di Scienze Fisiche, Univ. di Napoli ``Federico
II'', I-80126 Italy}

\email{cataudella@na.infn.it}

\begin{abstract}
Transport properties, spectral function and optical conductivity of the adiabatic one-dimensional Su-Schrieffer-Heeger (SSH) model are studied with particular emphasis on the model parameters suitable for Rubrene single crystals based field effect transistors. We show that the mobility, calculated by using the Kubo formula for conductivity, vanishes unless we introduce an "ad hoc" broadening of the system energy levels. Furthermore, the apparent contradiction between angle resolved photoemission data and transport properties is clarified by studying the behavior of the spectral function. Finally, a peak in the optical conductivity at very low energy is obtained and discussed in connection with the available experimental data for Rubrene based devices.
\end{abstract}
\maketitle

\section{introduction}
The organic field-effect transistors (OFET) play an important role in the field of the so-called Ôplastic electronicsÕ. Recently,  OFET, based on single crystals of oligoacene molecules, have been developed.\cite{hasegawa} From the study of these systems it stems that charge mobility in oligoacene single crystals can be one order of magnitude larger than that one with non ordered molecules.  Among them the more promising are those based on  Rubrene crystals that exhibit a strong anysotropy and the largest mobility measured in organic semiconductors.

In spite of the important technological impact of such devices, the intrinsic transport mechanism acting in rubrene is not fully understood. Actually, while the temperature dependence of the mobility exhibits a power-law ($\mu_e\sim T^{-\delta}$ with $\delta\simeq 2$)\cite{morpurgo,ostroverkhova,nature} reminding that one of mobile charge carriers (band transport), computational data indicate that the scattering length of the charge carrier becomes too short to be compatible with band transport.\cite{cheng}
Furthermore, some spectroscopic evidences support the localization of the charge carrier within one or few molecules\cite{troisiprl_12} and, on the other side, angle resolved photoemission spectroscopy (ARPES)\cite{arpes} seems to point again towards a band-like behavior of the charge carriers.   
 
In this framework a very interesting and simple one-dimensional model has been recently introduced by Troisi\cite{troisi_prl} where the charge carriers interact with the intermolecular modes leading to a modulation of the charge carrier hopping. The proposed model is somehow very close to the SSH model introduced in a different context.\cite{SSH} 

The model has been studied within the adiabatic limit (phonons obey a classical dynamics) in Ref.[\onlinecite{troisi_prl}] by using an approximated dynamical approach and the results have been, then, confirmed and extended by Ciuchi and Fratini\cite{fc} who used, instead, a thermodynamic approach where the vertex renormalizations are neglected. The main results claimed by those authors is that a power law for the mobility temperature dependence can be recovered within the proposed model and that the charge carriers involved in the transport undergo a "dynamical localization".  

From this brief discussion it is clear that a systematic study, "numerically" exact, of the transport properties of the SSH model in the adiabatic limit by using the Kubo formula for the conductivity is quite important. We will show that, at the thermodynamical equilibrium for both electrons and lattice degrees of freedom and including all the vertex rinormalizations, the mobility is dominated, as expected,\cite{anderson} by an "ad hoc" broadening of the energy levels. A proper choice of this broadening energy, taking into account in a qualitative way the missing energy scales in the adiabatic limit, is able to recover the power-law observed in the experiments. Then, we discuss in some detail how the coupling strength among charge carriers and phonons is able to modify temperature dependence of the mobility and charge carrier localization. We also measure the temperature dependent participation number showing a very weak temperature dependence in the experimental window and, on the contrary, a very strong dependence on the coupling strength between charge carriers and intermolecular phonons.

Furthermore, the analysis of the spectral function allows us to reconcile the results provided by the ARPES data (apparent band like description) with the computational observation that the scattering length of the charge carrier becomes too short to be compatible with band transport.\cite{cheng}

Finally we focus on the optical conductivity of the model emphasizing the dependence on temperature and charge carrier density. This allows us to individuate a low energy peak, below any charge transfer excitation,\cite{petrenko} that compares well with recent experimental results.\cite{li,fisher}

\section{The model}

The transport properties, the spectral function and optical conductivity of rubrene will be studied within the SSH model introduced by Troisi.\cite{troisi_prl} In this model the charge carriers move in a one-dimensional lattice hopping between next neighboring sites with a probability amplitude controlled by the relative position of the ions at the sites involved in the hopping. It can be summarized in the following model hamiltonian:     
\begin{equation}
H=\frac{m}{2}\sum_{i}\left(\dot{x}_{i}\right)^{2}+\frac{k}{2}\sum_{i}\left(x_{i}\right)^{2}
+H_{el}
\label{h}
\end{equation}
where
\begin{equation}
H_{el}=\sum_{i}\left[-t+\alpha\left(x_{i+1}-x_{i}\right)\right]\left(c_{i}^{\dagger}c_{i+1}+c_{i+1}^{\dagger}c_{i}\right).
\label{hel}
\end{equation}

In eqs. (\ref{h}) and (\ref{hel}) $t$ is the bare electron hopping, $\alpha$ is the coupling constant that controls the link between the electron hopping and ion displacement ($x_i$) and, finally, $m$ and $k$ are the mass and the elastic constant of ions, respectively. We emphasize that the electron dynamics is fully quantum ($c_{i}^{\dagger}$ being the charge carrier creation operator) while the ion dynamics is assumed classic. The latter approximation is well justified from the typical values of phonon frequencies $\omega_0$ and hopping constant $t$ for rubrene. Following Ref.[\onlinecite{troisi_am}] $\hbar\omega_0\simeq 6 meV$ and $t\simeq 140 meV$ leading to an adiabatic ratio $\gamma=\hbar\omega_0/t\simeq 0.04$. As we will discuss later, even if we are in a strong adiabatic regime the charge carrier mobility can be still affected by very small quantum effects due to the one-dimensional nature of the model. On the other hand, the finite frequency conductivity and other properties are not affected significantly by quantum fluctuations in this regime. In the following we will use dimensionless units measuring lengths in units of $l_0=(\hbar/(2m\omega_0))^{1/2}$ and energies in units of $t$.

Taking advantage of the classic nature of the lattice distortions, the partition function can be written as 

\begin{equation}
Z=\left(\frac{2m\pi}{\beta}\right)^{L/2}\sum_{\{x_{i}\}}\left\{ \exp\left[-\beta\frac{k}{2}\sum_{i}\left(x_{i}\right)^{2}\right]Z_{el}[\{x_{i}\}]\right\}
\label{Zeta}
\end{equation}
where $Z_{el}[\{x_{i}\}]$ is the quantum partition function of the electron subsystem given a deformation configuration $\left\{x_i\right\}$, $L$ is lattice size and $\beta=1/K_BT$.   

Following Michielsen and de Raedt \cite{dereadt} it is possible to estimate $Z$ by using a Monte Carlo approach for the classical degrees of freedom and exact diagonalization for the electron quantum dynamics. The method provides an approximation-free partition function of the model in the semiclassical limit. The only limitation is due to the computational time being controlled by the $L\times L$ matrix diagonalization. This constrains our analysis up to $L=128$. In order to reduce the size effect, we use periodic boundary conditions. 

Within the same framework it is also possible to calculate the spectral function and the optical conductivity averaging the electronic properties at a given ion displacement configuration over the entire set of  configurations weighted by Monte Carlo dynamics.\cite{dagotto} For instance, in the case of conductivity, for each configuration we calculate

\begin{eqnarray}
&&Re[\sigma(\omega;\{x_i\})]=\frac{(e a)^2}{\hbar}\frac{2\pi}{V\omega} \nonumber\\
&&\sum_{\lambda,\lambda'}\left(p_\lambda-p_{\lambda'}\right)\left|\left\langle \lambda|J|\lambda'\right\rangle\right|^2\delta(E_{\lambda}-E_{\lambda'}+\omega)
\label{sigma_mu}\\\nonumber
\end{eqnarray}
where

\begin{equation}
 p_\lambda=\frac{1}{\exp[\beta (E_\lambda-\mu)]+1} .
\label{fermi}
\end{equation}
In eq.(\ref{sigma_mu}) $V=L a$ is the system volume ($a$ is the distance between next neighboring sites) and in eq.(\ref{fermi}) $\mu$ is the chemical potential, while $E_\lambda$ are the eigenvalues.

The matrix element in eq.(\ref{sigma_mu}) can be expressed in terms of the eigenstates of $H_{el}$. By using the unitary matrix that diagonalizes $H_{el}$, $U(i,\lambda)$, we can write
\begin{eqnarray}
\left\langle \lambda|J|\lambda'\right\rangle&=&\sum_i t_i \sum_{\mu,\mu'}\left[ U(i,\mu)U(i+1,\mu')\left\langle \lambda|c_{\mu}^{\dagger}c_{\mu'}|\lambda'\right\rangle\right. \nonumber \\
&-&\left. U(i+1,\mu)U(i,\mu')\left\langle \lambda|c_{\mu}^{\dagger}c_{\mu'}|\lambda' \right\rangle \right]\\
&=&\sum_i  t_i \left[ U(i,\lambda)U(i+1,\lambda')-U(i+1,\lambda)U(i,\lambda')\right]\nonumber\\
\end{eqnarray}
where
\[t_i=-t+\alpha\left(x_{i+1}-x_{i}\right).\]

Then, the mobility can be defined as:
 \begin{equation}
\mu_e=\frac{1}{\rho e}\lim_{\omega\rightarrow 0^+}Re[\sigma(\omega)]
\label{mobility_mu}
\end{equation}
where $\sigma(\omega)$ is obtained by averaging over the displacement configurations $\sigma(\omega;\{x_i\})$ and the density, $\rho=N_e/L$, ($N_e$ being the charge carrier number) is
\begin{equation}
\rho=2\sum_{n}p_n.
\label{density}
\end{equation}

In the following we will focus our attention mainly to the limiting case in which a single electron is present in the system. Then $p_\lambda\mapsto \exp[-\beta E_\lambda]/Z_{el}$ in eq.(\ref{sigma_mu}) and the factor two, due to the spin degeneracy, drops out.

\section{Temperature dependence of mobility}

As mentioned in the introduction, the mobility at the thermodynamic equilibrium, calculated assuming that the ion displacements are classical variables, is dominated by the energy broadening that we have to include in eq.(\ref{sigma_mu}). Indeed, in order to use that expression for any finite lattice we have to replace the delta function with a Lorentzian
\begin{equation}
\frac{1}{\pi}\delta(E_{\lambda}-E_{\lambda'}+\omega)\mapsto\frac{\eta}{(E_{\lambda}-E_{\lambda'}+\omega)^2+\eta^2}.
\end{equation}

The correct expression is obtained, then, for $L\mapsto\infty$ and $\eta\mapsto 0$. It can be easily shown that, as expected, this procedure leads to a vanishing mobility. Indeed, in the adiabatic limit, our calculation is equivalent to the classical problem of a particle in presence of off-diagonal disorder that is characterized by a vanishing mobility and localized eigen-functions except for that one corresponding to the zero energy eigenvalue.\cite{cohen,economou} It is then clear that the system will exhibit a finite mobility only if we include a broadening representing the small (relevant) energy scale not included in the original model. In our opinion the most important missing energy scale is the quantum phonon ground state energy
$\hbar\omega_0/2$. The effect of this energy scale is usually negligible in the adiabatic limit, but becomes relevant in 1D systems for the optical conductivity at energies less than $\hbar\omega_0/2$ and, then, it is crucial for the mobility. 

\begin{figure}[!h]
\flushleft
\includegraphics[scale=0.9]{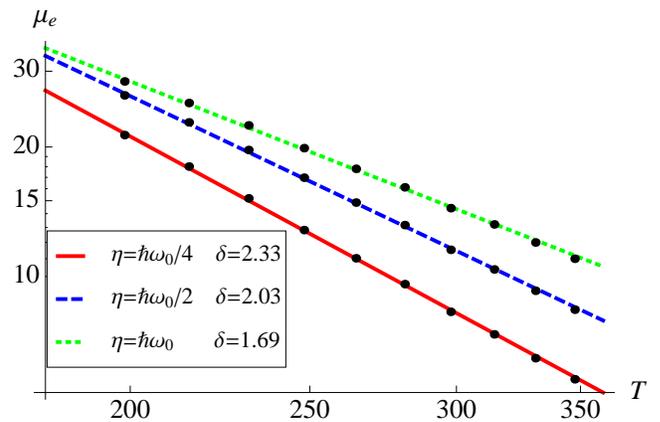}
\caption{Mobility vs Temperature for different values of $\eta$. Mobility is in units of $cm^{-2}/(s\cdot Volt)$ and the temperature in Kelvin.  The system size is $L=64$. }
\label{mobility_1}
\end{figure}

In Fig.(\ref{mobility_1}) we show the temperature dependence of the mobility for the model parameters suitable for rubrene:\cite{troisi_am} $\alpha/t=0.09$, $\omega_0/t=0.04$. We restrict ourselves to the case of a single charge carrier and plot the mobility for different values of $\eta$. It comes out that at $\eta=\hbar\omega_0/2$ a power-law behavior $\mu_e\sim T^{-\delta}$ with $\delta=2.03$ is recovered. On the other hand, a smaller (larger) value of $\eta$ provides a larger (smaller) exponent. In our opinion this is a strong indication that quantum fluctuations are indeed important in this system.  It is also worth noticing that, at the thermodynamic equilibrium, the mobility does not depend on the ion mass [see eq.(\ref{Zeta})]. Both the results, the need of an "external" energy scale and the mass independence of mobility, are not recovered within the dynamical approach proposed in Ref.[\onlinecite{troisi_prl}] pointing out that the dynamical and the equilibrium thermodynamic approaches are not equivalent. We also note that the mobility values we find are larger than those obtained in Ref.[\onlinecite{troisi_prl}] and surprisingly close to the experimental values.\cite{hasegawa} 

\begin{figure}[!h]
\centering
\includegraphics[scale=0.8]{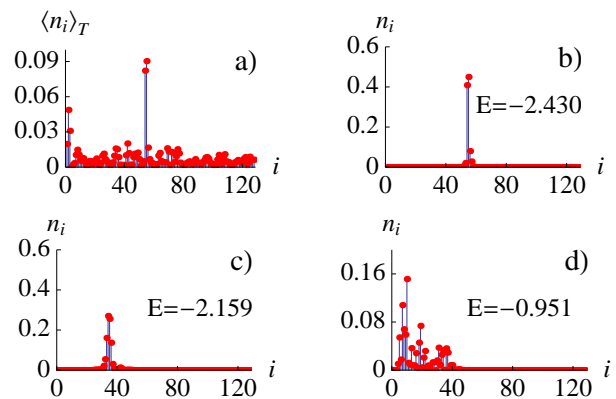}
\caption{Snapshot of the electron wave-function at (typical) given deformation configuration at $T=347K$. Panel a): average density versus lattice sites; panel b)-d): average density at fixed electron energy versus lattice sites. The system size is $L=128$ and the energy $E$ is measured in units of $t$.}
\label{snapshot}
\end{figure}

Summarizing, we have shown that the model of eq.(\ref{h}) in the adiabatic limit and at the thermodynamic equilibrium is able to recover the power-law observed in the experiments [see Fig.(\ref{mobility_1})] only introducing an "ad hoc" energy broadening that we associate to the quantum lattice fluctuations. On the other hand our analysis allows us to give a simple explanation of the physical mechanism responsible for the finite mobility. Due to the one dimensional nature of the model and the assumption that the lattice oscillations are classical variables all the  charge carrier wave-functions are localized [see the snapshot reported in Fig.(\ref{snapshot})]. In particular, as already proven in Ref.[\onlinecite{fc}], a more detailed analysis shows that the wave functions whose energies are closer to band border are even more localized than those well within the energy band. The only exception to this description is given by the wave-function corresponding to vanishing energies that exhibits an anomalous behavior.\cite{cohen,economou}

This scenario leads to a vanishing mobility unless we assume an energy broadening that is able to provide the needed eigenvalue overlap. 
The power-law observed experimentally in the temperature range $2\hbar\omega_0<K_BT<5\hbar\omega_0$ is, then, the result of entirely incoherent processes that have nothing to do with the band transport. It is worth noticing that this temperature behavior is due to the very small phonon energy involved in the scattering and it corresponds, in simple metals, to that typical of very high temperature regime.\cite{gunnarsson, millis}  Finally, we emphasize that only a full quantum analysis will clarify how this scenario is modified at lower temperatures where the adiabatic approximation breaks down and the quantum fluctuations enter the problem in a more intrinsic way giving rise also to a Boltzmann-like contribution of the charge carriers that cannot be recovered in the present adiabatic approximation. 

For a better understanding of the physical origin of the temperature dependence of the mobility we note that, as first observed by Troisi\cite{troisi_prl} and discussed in more detail by Fratini and Ciuchi,\cite{fc} the wave-function localization increases with temperature. In the equilibrium thermodynamic approach used in this paper the localization stems from the analogy with the disorder problem: increasing the temperature is equivalent to strengthen the disorder. On the other hand, if we measure the thermal average of any physical quantity the temperature enters not only through the distribution of ion displacements (as in the equivalent disorder problem), but also in the thermal average of the physical quantity of interest at any fixed lattice configuration. Actually, with increasing the temperature, the thermal average involves more and more wave-functions corresponding to larger and larger eigenvalues which correspond to {\em less} localized wave-functions. The competition between the two effects can give non trivial results. In order to clarify this issue we have calculated the temperature dependent participation number that provides a measure of the average effective localization:
\begin{equation}
P=\left[\sum_{i=1}^N \langle n_i \rangle^2 \right]^{-1}.
\end{equation}
$P$ ranges from the number of the lattice sites ($N$)  for delocalized states (traslational invariance) to $1$ for a state fully localized on a single site. 

\begin{figure}[!h]
\centering
\includegraphics[scale=0.9]{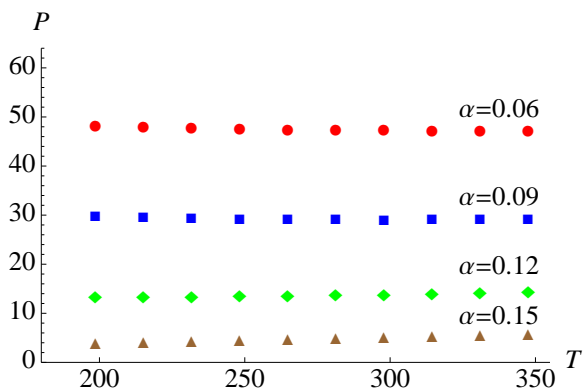}
\caption{Participation number, $P$, as function of temperature $T$ for different values of the coupling constant $\alpha$. The temperature, $T$, is measured in Kelvin and the system size is $L=64$.}
\label{participation}
\end{figure}

In Fig.(\ref{participation}) we report the temperature behavior of $P$ showing that, in the temperature regime of interest, $P$ does not exhibit a significant change. In this sense we think that the fact that the mobility decrease with the temperature is not strictly related to the wave-function localization, but rather it is due to an increase of the scattering rate among the electron and the lattice deformation. 

On the other hand, it is interesting to emphasize the strong dependence of the participation number, $P$, with the coupling constant $\alpha$. In particular for $\alpha=0.15$ $P$ becomes equal to few lattice sites. At this value of the coupling constant the nature of the ground state has changed: the electron form a bond polaron\cite{grilli, lamagna, zoli, noi_perroni, zhao, noi} that is characterized by a very large effective mass.

\begin{figure}[h]
\centering
\includegraphics[scale=0.9]{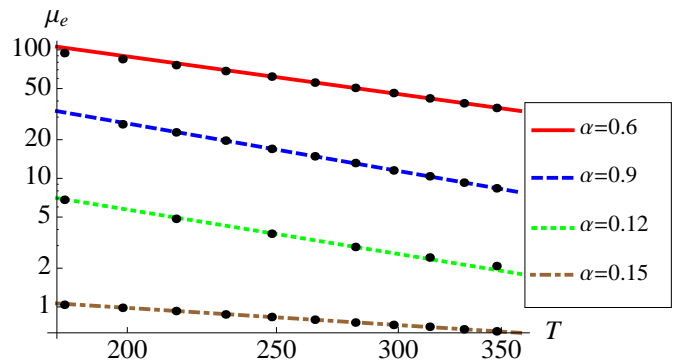}
\caption{mobility, $\mu_e$, as function of temperature $T$ for different values of the coupling constant $\alpha$. Mobility is in units of $cm^{-2}/(s\cdot Volt)$ and the temperature in Kelvin. The system size is $L=64$.}
\label{m_alfa}
\end{figure}

We end up this section presenting the mobility for different coupling constants [Fig.(\ref{m_alfa})]. We observe that the power-law behavior is very robust and is recovered even in the most localized case reinforcing the idea that it is a purely incoherent mechanism. Finally we emphasize that, while the absolute value of the mobility decreases, the power $\delta$ becomes smaller signaling a rather complex link between mobility and localization.

\section{Spectral function}
\begin{figure}[!h]
\centering
\includegraphics[scale=0.9]{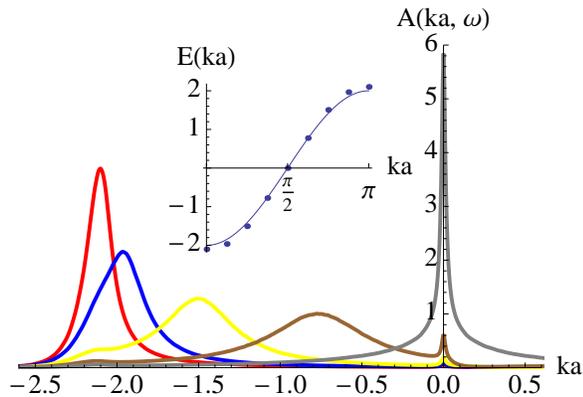}
\caption{The spectral function for different values of the wave-vector $k$. Measuring $k$ in units of the inverse of the lattice spacing, $a$, we show from left to right the values $ka=0,\frac{\pi}{8},\frac{\pi}{4},\frac{3\pi}{8},\frac{\pi}{2}$. In the inset we report the effective electron band (points) compared with the bare electronic band (full line). The temperature $K_BT/t=0.21$ and $L=64$.}
\label{Ak}
\end{figure}

In the previous section we have shown that the equilibrium mobility in our model is due to purely incoherent processes and it cannot be ascribed to a simple band-like description. On the other hand, as mentioned in the introduction, ARPES measurements\cite{arpes} show that the effective electronic energy dispersion extracted from the $k-$dependent spectral function, $A(k,\omega)$, is very close to a simple $\cos(ka)$ band. This finding is usually considered as an indication in favor of a band-like scenario. In order to clarify this apparent contradiction and with the aim to validate the model studied in the present paper, we have calculated the spectral function for different $k$ values and extracted the effective electronic band (see Fig.(\ref{Ak})). 

As it can be seen in the inset, the energy dispersion obtained following the main maxima of the spectral function is very close to the bare band in agreement with the ARPES. However, a more careful analysis shows that the peaks exhibited by the spectral function are very broad\cite{nota1} making the quasi-particle description not well founded. Actually, for $k=\pi/4$ is evident a double peak structure that persists for lower $k$ values even if in a less evident way. Finally, as expected, the spectral function presents an anomalous behavior at $E=0$ where the associated wave-function is not localized.\cite{cohen,economou} In our opinion, the present analysis provided a simple explanation of the apparent contradiction between ARPES and transport measurements: the absence of well defined quasi-particles makes the band-like description not applicable but, at the same time, the main peak of the spectral function inherits the bare band dispersion. Finally, we would mention that the effective energy dispersion depends very little with temperature in the range analyzed in this paper showing a very little increase of the bandwidth. In this sense the system is not characterized by any quasi-particle with heavy effective mass.

\section{Optical conductivity}

\begin{figure}[h]
\centering
\includegraphics[scale=0.9]{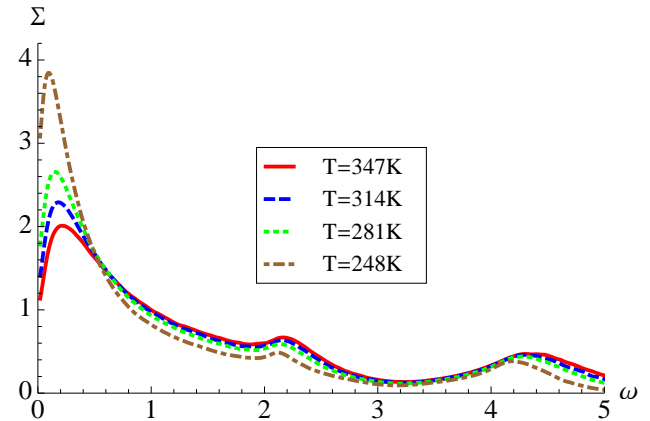}
\caption{The dimensional optical conductivity, $\Sigma=\sigma(\omega)\hbar V /[2\pi (ea)^2]$, as a function of the dimensionless photon energy $\omega=\hbar\Omega/t$  for different temperatures (measured in Kelvins). The lattice size is $L=64$.}
\label{OC_09}
\end{figure}

As mentioned in the introduction, measurements of the optical conductivity (OC) are an important tool to investigate the properties of OFET devices.\cite{fisher,li} For this reason we have calculated OC within the studied model. In Fig.(\ref{OC_09}) we show the OC for the parameter values relevant.\cite{troisi_am} The OC exhibits a clear peak at low energies $\hbar\omega\simeq 0.2t$ whose intensity decreases with the temperature moving slightly towards high energies. It is worth noticing that, unlike the mobility, the peak position does not depend on the broadening energy that we still choose equal to $\hbar\omega_0/2$. The result is of some interest since there are experimental evidence\cite{fisher,li} that, indeed, a peak is present at energies about $62meV$  ($500cm^{-1}$) lower than any charge transfer process.\cite{petrenko} Assuming $t\simeq140meV$ our estimate is a factor two lower than the measured value. Our estimation has to be considered quite reasonable for the very simple one-dimensional model we adopted. Actually, as we will show in the following, our estimate can be even improved modifying slightly the phenomenological parameters of the model.

We observe that the OC of the model exhibits also smaller structure at higher energies. However, for energy very much larger than the bare charge carrier hopping, $t$, the model cannot be trusted in the framework of the rubrene OFET since at higher energies many charge transfer processes are observed\cite{petrenko} that are not taken into account in the model presented here. Nevertheless these higher energies structures are still of interest for the model itself that has been proposed in many contexts. 
   
\begin{figure}[h]
\centering
\includegraphics[scale=0.9]{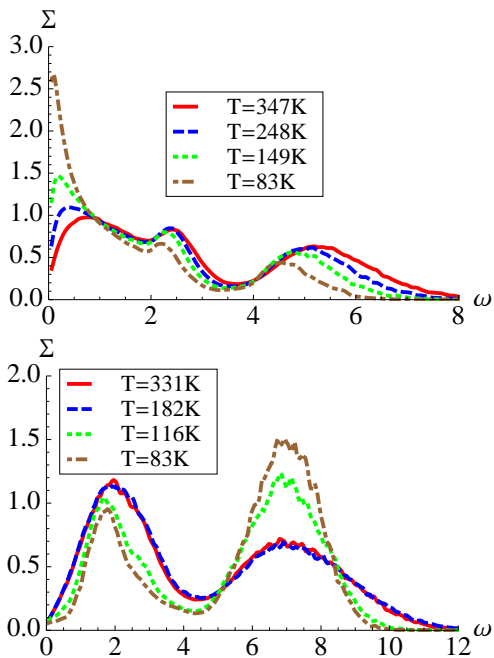}
\caption{The dimensionless optical conductivity, $\Sigma=\sigma(\omega)\hbar V /[2\pi (ea)^2]$, as a function of the dimensionless photon energy $\omega=\hbar\Omega/t$  for different temperature (measured in Kelvins). The lattice size is $L=64$. Upper panel $\alpha=0.12$, lower panel $\alpha=0.15$}
\label{OC_12_15}
\end{figure}

As in the case of the mobility, it is interesting to study how the OC changes with the strength of the coupling constant, $\alpha$, which, as discussed in the previous section, makes the system more and more "localized".  As shown in Fig.(\ref{OC_12_15}), when $\alpha$ increases, all the spectral weight move towards higher energies. In particular, at $\alpha=0.12$ [Fig.(\ref{OC_12_15}a], the low energy peak move to $\omega\simeq0.5t$, becomes broader and looses intensity. This analysis suggests that a larger value of the low energy peak can be obtained by tuning the value of the charge-lattice coupling providing a better agreement with the experimental data. A further increase of $\alpha$ drives the system towards a localized state associated with a very large increase of the effective mass and a clear optical gap opens up at low energies signaling the bond polaron formation ($\alpha=0.15$). 

\section{Transport properties and optical conductivity at finite density}
Up to now we have focused our attention on the case of a single particle interacting with the lattice fluctuations, but the experiments, of course, are performed  at a finite (even if small) particle density. Can we expect significant differences?
The approach used in this work has the advantage that can deal very naturally with a finite number of particle and, then, we can address this point. 
In Fig.(\ref{OC_12_15}) we show the mobility as a function of the charge carrier density, $\rho$ for low densities ($\rho< 0.12$) and different temperatures. We restrict our analysis to small density because OFET are characterized by extremely low density and also since we completely neglect charge-charge interaction in the model studied. It is clear that the mobility is significantly affected by the carrier density reducing its value when $\rho$  increases. The single particle case studied in the previous sections provides, then, the maximum value for the mobility in this model. The non uniform behavior with temperature also signals that the temperature exponent of the mobility changes with the density. From our analysis stems out that, although the density increase provides an obvious increase of the conductivity (it is proportional to $\rho$), the mobility decrease compensates such increase and can even cancel it. Therefore, we expect a less linear increase of the conductivity with density. We interpret this result as the effect of the charge-charge effective interaction (mediated by the lattice fluctuations) that represents a further scattering mechanism for the charge carriers. The sensitivity of the mobility to the effective interaction suggests that the inclusion of direct Coulomb interaction among the charges could be important even at low densities.
\begin{figure}[h]
\centering
\includegraphics[scale=0.8]{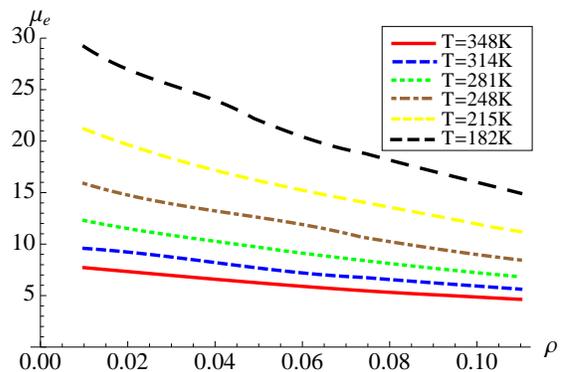}
\caption{Mobility vs. charge carrier density, $\rho$ for different temperature
[$T=348K$ full line, $T=314K$ dashed line, $T=281K$ dotted line, $T=248K$ dashed-dotted line, $T=215K$ long dashed line, $T=182K$ long-long dashed line]. The lattice size is $L=64$.}
\label{mu_rho}
\end{figure}

\begin{figure}[h]
\centering
\includegraphics[scale=0.8]{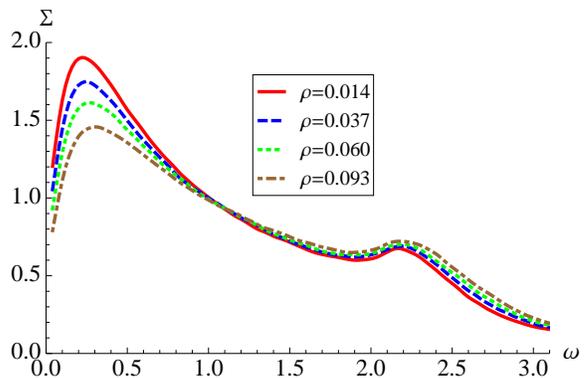}
\caption{Optical conductivity, $\Sigma=\sigma(\omega)\hbar V /[2\pi (ea)^2]$, vs. dimensionless photon energy, $\omega=\hbar\Omega/t$  for different charge carrier density $\rho$ [$\rho=0.014$ full line, $\rho=0.014$ dashed line, $\rho=0.014$ dotted line, $\rho=0.014$ dashed-dotted line]. The temperature is $T=348K$. Energy are measured in units of the bare hopping, $t$.}
\label{mu_OC}
\end{figure}

We end up this section presenting the OC at different densities for the model parameters appropriate for rubrene. We still get a low energy peak as in the case of a single particle, but the peak position moves towards higher energies and the intensity decreases[Fig.(\ref{mu_rho})] increasing the density $\rho$. This behavior, as it has been shown in Fig.(\ref{mu_OC}), is accompanied by a decrease of the mobility. We note that the observed behavior at finite density reminds that obtained increasing the coupling $\alpha$ [see Fig.(\ref{OC_12_15})].

\section{Conclusions}     
In this work we have studied the thermodynamic equilibrium properties of the SSH model in 1D assuming the ion displacements as classical variables. We focused our attention on mobility, spectral function and optical conductivity in the adiabatic regime ($\gamma=0.04$). Actually, this regime is particularly interesting for its connection with the low energy physics of OFET based on Rubrene single crystal as proposed  in Refs.\onlinecite{troisi_prl,troisi_am}. We find that, as expected at the thermodynamic equilibrium, the model mobility vanishes unless an "ad hoc" energy broadening is introduced in the model. Interestingly if we choose this energy broadening of the order of $\hbar\omega_0/2$ we are able to recover the temperature dependence observed in the experiments. This result suggests that the lattice quantum fluctuations not considered in our approach are crucial for the mobility even in a strong adiabatic regime. It is worth noticing that the present approach shows a significant difference with the approximated dynamical approach\cite{troisi_prl,troisi_am} where a finite mobility is obtained without invoking an "ad hoc" energy broadening. In our opinion the difference stems from the fact that the dynamics adopted does not bring the system to the full thermodynamic equilibrium. 

We also showed that the single-particle spectral function, $A(k,\omega)$ exhibits very broad peaks that cannot be described as a simple quasi-particle. However, the position of the main peak inherits the bare electron energy dispersion. This result could explain the apparent contradiction between ARPES and the power law temperature of the mobility observed in the transport measurement of Rubrene OFET.

Finally, the analysis of the OC of the model suggests the existence of a low energy peak in the energy range $0.2t<\omega<0.6t$ depending on the charge-lattice coupling $\alpha$ and the charge carrier density. As discussed in the previous sections, this result is of some interest since there are experimental evidence on single crystal Rubrene OFET\cite{fisher,li} showing that, indeed, a peak is present at energies about $62meV$ lower than any charge transfer process.\cite{petrenko} Assuming in our model a bare charge hopping, $t\simeq140meV$, our estimate is in a reasonable agreement with the measured value. More experimental data, in particular temperature dependence of the peak,  and more realistic models will be needed for a full characterization of this low energy OC peak.         

\section{Acknowledgements}
We thank Sergio Ciuchi and Simone Fratini for useful and stimulating discussions. This work was partially supported by MIUR-PRIN 2007 under Prot. No. 2207FW3MJC003 and University of Napoli "Federico II" under grant FARO 2010.


\begin{thebibliography}{99}
\bibitem{hasegawa}T. Hasegawa and J. Takeya, Sci. Technol. Adv. Mater. {\bf 10}, 24314 (2009).
\bibitem{morpurgo}M. E. Gershenson, V. Podzorov and A. F. Morpurgo, Rev. Mod. Phys. {\bf 78}, 973 (2006).
\bibitem{ostroverkhova}O. Ostroverkhova et al., Phys. Rev. B {\bf 71}, 035204 (2005).
\bibitem{nature}I.N. Hulea, S. Fratini, H. Xie, C.L. Mulder, N.N. Iossad, G. Rastelli, S. Ciuchi, and A.F. Morpurgo, Nature Materials 5, 982 (2006).
\bibitem{cheng}Y. C. Cheng et al., J. Chem. Phys. {\bf 118}, 3764 (2003).
\bibitem{troisiprl_12} P. J. Brown, H. Sirringhaus, M. Harrison, M. Shkunov, R.H. Friend, Phys. Rev. B {\bf 63}, 125204 (2001).
\bibitem{arpes} Huanjun Ding, Colin Reese, Antti J. MŠkinen, Zhenan Bao, and Yongli Gao, App. Phys. Lett. {\bf 96}, 222106 (2010).
\bibitem{troisi_prl} A. Troisi and G. Orlandi, Phys. Rev. Lett.{\bf 96}, 086601 (2006).
\bibitem{SSH}W. P. Su, J. R. Schrieffer, and A. J. Heeger, Phys. Rev. Lett. {\bf 42},1698 (1979).
\bibitem{fc}S. Fratini and S. Ciuchi, Phys. Rev. Lett. {\bf103}, 266601 (2009).
\bibitem{anderson}P. W. Anderson, Phys. Rev. {\bf 109}, 1492 (1958).
\bibitem{petrenko}T. Petrenko, O. Krylova, F. Neese and M. Sokolowski, New Journal of Physics {\bf 11}, 15001 (2009).
\bibitem{li}Z. Q. Li, V. Podzorov, N. Sai, M. C. Martin, M. E. Gershenson, M. Di Ventra, and D. N. Basov, Phys. Rev. Lett. {\bf 99}, 016403 (2007).
\bibitem{fisher}M. Fischer, M. Dressel, B. Gompf A. K. Tripathi and J. Pflaum, Appl. Phys. Lett. {\bf 89}, 182103 (2006).
\bibitem{kakuta}H. Kakuta, T. Hirahara, I. Matsuda, T. Nagao, S. Hasegawa, N. Ueno, and K. Sakamoto,  Phys. Rev. Lett. {\bf 98}, 247601 (2007). 
\bibitem{troisi_am} A. Troisi, Adv. Mater. {\bf 19}, 2000 (2007).
\bibitem{dereadt}K. Michielsen and H. de Raedt, Modern Physics Letters B{\bf 10}, 467 (1996).
\bibitem{dagotto}E. Dagotto, T. Hotta, A. Moreo, Physics Reports 344, 153 (2001).
\bibitem{cohen}G. Theodorou and M.H. Cohen, Phys. Rev. B {\bf 13}, 4597 (1976).
\bibitem{economou}C.M. Soukoulis and E.N. Economou, Phys. Rev. B {\bf 24}, 5698 (1981); L. Fleishman and D.C. Licciardello, J. Phys. C{\bf 10}, L125 (1977).
\bibitem{millis}A. J. Millis, Jun Hu and S. Das Sarma, Phys. Rev. Lett. {\bf 82}, 2354 (1999).
\bibitem{gunnarsson}O. Gunnarson, M. Calandra and J.E. Han, Rev. of Mod. Phys. {\bf 73}, (2003).
\bibitem{grilli}M. Capone, W. Stephan, and M. Grilli, Phys. Rev. B {\bf 56}, 4484  (1997).
\bibitem{lamagna}A. La Magna and R. Pucci, Phys. Rev. B {\bf 55}, 6296 (1997).
\bibitem{zoli}M. Zoli, Phys. Rev. B {\bf 66}, 012303 (2002); M. Zoli, "Path Integral Methods in the SSH Polaron model", in \textit{Polarons in Advanced Materials} ed. by A.S. Alexandrov, Springer (2007), p. 231. 
\bibitem{noi_perroni}C. A. Perroni, E. Piegari, M. Capone and V. Cataudella, Phys. Rev. B {\bf 69}, 174301 (2004). 
\bibitem{zhao}Yang Zhao, Guangqi Li, Jin Sun, and Weihua wang, J. of Chem. Phys. {\bf 129},124114 (2008).
\bibitem{noi} D. Marchand, G. De Filippis, V. Cataudella, M. Berciu, N. Nagaosa
N. V. ProkofÕev, P. C. E. Stamp and A. S. Mishchenko, to be published.
\bibitem{nota1} The results shown in Fig.(\ref{Ak}) do not depend on the broadening energy $\eta$, except for the zero energy peak whose width is actually $\eta$.


\end{thebibliography}
\end{document}